\documentclass[12pt]{article}
\usepackage{amsmath,float,amssymb,amsthm,amsxtra,overpic,bbm,bm,epsfig}
\def\thefootnote{\fnsymbol{footnote}}

\begin{document}

\vspace{0.2cm}

\begin{center}
{\large\bf Terrestrial matter effects on reactor antineutrino
oscillations \\ at JUNO or RENO-50: how small is small?}
\end{center}

\vspace{0.1cm}

\begin{center}
{\bf Yu-Feng Li}\footnote{E-mail: liyufeng@ihep.ac.cn}, ~ {\bf
Yifang Wang}\footnote{E-mail: yfwang@ihep.ac.cn}, ~
{\bf Zhi-zhong Xing}\footnote{E-mail: xingzz@ihep.ac.cn} \\
{\small Institute of High Energy Physics, Chinese Academy of
Sciences, Beijing 100049, China \\
and \\
School of Physical Sciences, University of Chinese Academy of
Sciences, Beijing 100049, China}
\end{center}

\vspace{1.5cm}

\begin{abstract}
We have carefully examined, in both analytical and numerical ways,
how small the terrestrial matter effects can be in a given
medium-baseline reactor antineutrino oscillation experiment like
JUNO or RENO-50. Taking the ongoing JUNO experiment for example, we
show that the inclusion of terrestrial matter effects may reduce the
sensitivity of the neutrino mass ordering measurement by $\Delta
\chi^2_{\rm MO}\simeq 0.6$, and a neglect of such effects may shift
the best-fit values of the flavor mixing angle $\theta^{}_{12}$ and
the neutrino mass-squared difference $\Delta^{}_{21}$ by about
$1\sigma$ to $2\sigma$ in the future data analysis. In addition, a
preliminary estimate indicates that a $2\sigma$ sensitivity of
establishing the terrestrial matter effects can be achieved for
about 10 years of data taking at JUNO with the help of a proper near
detector implementation.
\end{abstract}

\begin{flushleft}
\hspace{0.8cm} PACS number(s): 14.60.Pq, 13.10.+q, 25.30.Pt \\
\hspace{0.8cm} Keywords:  terrestrial matter effects, reactor
antineutrino oscillations
\end{flushleft}

\def\thefootnote{\arabic{footnote}}
\setcounter{footnote}{0}

\vspace{2cm}


The approved JUNO project in China is a flagship of the
new-generation medium-baseline reactor antineutrino oscillation
experiments~\cite{JUNO,JUNOCDR}, and its primary physics target is
to probe the intriguing neutrino mass ordering~\cite{MHrev1,MHrev2}
(i.e., whether $m^{}_1 < m^{}_2 < m^{}_3$ or $m^{}_3 < m^{}_1 <
m^{}_2$). A similar project in South Korea, the RENO-50
experiment~\cite{RENO}, has been proposed for the same purpose.
Since the typical energies of electron antineutrinos produced from a
reactor are around $4$ MeV, terrestrial matter effects are expected
to be negligibly small in a given $\overline{\nu}^{}_e \to
\overline{\nu}^{}_e$ oscillation experiment. However, a careful
examination of the sensitivity of measuring the neutrino mass
ordering to the matter-induced contamination has been lacking,
although some preliminary estimates of the matter effects on the
leptonic flavor mixing angles and neutrino mass-squared differences
have been made in this connection~\cite{Lisi1,Li-JUNO,Lisi2}.

In the present work we aim to evaluate how small the terrestrial
matter effects are and whether they can more or less affect the
precision measurements to be done in the JUNO and RENO-50
experiments. Our main results will be presented both numerically and
in some useful and instructive analytical approximations. A
remarkable observation is that the terrestrial matter contamination
may give rise to a correction close to $1\%$ to the quantity
associated with a crucial judgement of whether the neutrino mass
ordering is normal or inverted. Taking the ongoing JUNO experiment
as an example, we show that the inclusion of terrestrial matter
effects may reduce the sensitivity of the neutrino mass ordering
measurement by $\Delta \chi^2_{\rm MO}\simeq 0.6$, and a neglect of
such effects may shift the best-fit values of the flavor mixing
angle $\theta^{}_{12}$ and the neutrino mass-squared difference
$\Delta^{}_{21}$ by about $1\sigma$ to $2\sigma$ in the future data
analysis. Moreover, a preliminary estimate indicates that a
$2\sigma$ sensitivity of establishing the terrestrial matter effects
can be achieved for about 10 years of data taking at JUNO with the
help of a proper near detector implementation.

\vspace{0.4cm}

Let us begin with the effective Hamiltonian that is responsible for the
propagation of {\it antineutrinos} in matter~\cite{MSW,MSW2}
\begin{eqnarray}
\widetilde{\cal H}^{}_{\rm eff} = \frac{1}{2 E} \left[\widetilde{U}
\begin{pmatrix} \widetilde{m}^2_1 & 0 & 0 \cr 0 & \widetilde{m}^2_2
& 0 \cr 0 & 0 & \widetilde{m}^2_3 \cr \end{pmatrix}
\widetilde{U}^\dagger \right] = \frac{1}{2 E} \left[U
\begin{pmatrix} m^2_1 & 0 & 0 \cr 0 & m^2_2 & 0 \cr 0 & 0 & m^2_3
\cr \end{pmatrix} U^\dagger - \begin{pmatrix} A & 0 & 0 \cr 0 & 0 &
0 \cr 0 & 0 & 0 \cr \end{pmatrix} \right] \; ,
\end{eqnarray}
where $\widetilde{U}$ (or $U$) and $\widetilde{m}^{}_i$ (or
$m^{}_i$) stand respectively for the effective (or fundamental)
lepton flavor mixing matrix and neutrino masses in matter (or in
vacuum), and $A = 2\sqrt{2} \ G^{}_{\rm F} N^{}_e E$ with $G^{}_{\rm
F}$ being the Fermi constant and $N^{}_e$ being the background
density of electrons. In fact, $A$ itself and the minus sign in
front of $A$ denote the charged-current contribution to the coherent
$\overline{\nu}^{}_e e^-$ forward scattering in matter. Given a
constant matter profile which is a good approximation for the
reactor-based antineutrino oscillation experiments, one may
establish the exact analytical relations between $|U^{}_{e i}|^2$
and $|\widetilde{U}^{}_{e i}|^2$ as follows~\cite{Zhang}:
\begin{eqnarray}
|\widetilde{U}^{}_{e 1}|^2 \hspace{-0.15cm} & = & \hspace{-0.15cm}
\displaystyle +\frac{\Delta^\prime_{21} \Delta^\prime_{31}}
{\widetilde{\Delta}^{}_{21} \widetilde{\Delta}^{}_{31}}
|U^{}_{e 1}|^2 + \frac{\Delta^\prime_{11} \Delta^\prime_{31}}
{\widetilde{\Delta}^{}_{21} \widetilde{\Delta}^{}_{31}}
|U^{}_{e 2}|^2 + \frac{\Delta^\prime_{11} \Delta^\prime_{21}}
{\widetilde{\Delta}^{}_{21} \widetilde{\Delta}^{}_{31}}
|U^{}_{e 3}|^2 \; ,
\nonumber \\
|\widetilde{U}^{}_{e 2}|^2 \hspace{-0.15cm} & = & \hspace{-0.15cm}
\displaystyle -\frac{\Delta^\prime_{22} \Delta^\prime_{32}}
{\widetilde{\Delta}^{}_{21} \widetilde{\Delta}^{}_{32}}
|U^{}_{e 1}|^2 - \frac{\Delta^\prime_{12} \Delta^\prime_{32}}
{\widetilde{\Delta}^{}_{21} \widetilde{\Delta}^{}_{32}}
|U^{}_{e 2}|^2 - \frac{\Delta^\prime_{12} \Delta^\prime_{22}}
{\widetilde{\Delta}^{}_{21} \widetilde{\Delta}^{}_{32}}
|U^{}_{e 3}|^2 \; ,
\nonumber \\
|\widetilde{U}^{}_{e 3}|^2 \hspace{-0.15cm} & = & \hspace{-0.15cm}
\displaystyle +\frac{\Delta^\prime_{23} \Delta^\prime_{33}}
{\widetilde{\Delta}^{}_{31} \widetilde{\Delta}^{}_{32}}
|U^{}_{e 1}|^2 + \frac{\Delta^\prime_{13} \Delta^\prime_{33}}
{\widetilde{\Delta}^{}_{31} \widetilde{\Delta}^{}_{32}}
|U^{}_{e 2}|^2 + \frac{\Delta^\prime_{13} \Delta^\prime_{23}}
{\widetilde{\Delta}^{}_{31} \widetilde{\Delta}^{}_{32}}
|U^{}_{e 3}|^2 \; ,
\end{eqnarray}
where $\widetilde{\Delta}^{}_{ij} \equiv \widetilde{m}^2_i -
\widetilde{m}^2_j$ and $\Delta^\prime_{ij} \equiv m^2_i -
\widetilde{m}^2_j$ as compared with the fundamental neutrino
mass-squared differences $\Delta^{}_{ij} \equiv m^2_i - m^2_j$ (for
$i, j = 1, 2, 3$). To see the matter effects hidden in
$\widetilde{\Delta}^{}_{ij}$ and $\Delta^\prime_{ij}$ in a
transparent way, we take into account their approximate expressions
expanded in terms of two small parameters $\alpha \equiv
\Delta^{}_{21}/\Delta^{}_{31}$ and $\beta \equiv A/\Delta^{}_{31}$
in the normal neutrino mass ordering (i.e., $\Delta^{}_{31} >0$)
case \cite{XingZhu}:
\begin{eqnarray}
\widetilde{\Delta}^{}_{21} \hspace{-0.15cm} & \simeq &
\hspace{-0.15cm} \Delta^{}_{31} \left(1 -
\frac{3}{2} |U^{}_{e 3}|^2 \beta \right) \epsilon \; ,
\nonumber \\
\widetilde{\Delta}^{}_{31} \hspace{-0.15cm} & \simeq &
\hspace{-0.15cm} \Delta^{}_{31} \left[1 - \frac{1}{2} \alpha +
\frac{1}{2} \left(1 - 3|U^{}_{e 3}|^2 \right) \beta + \frac{1}{2}
\epsilon + \frac{3}{4} |U^{}_{e 3}|^2 \beta \left(2\beta -\epsilon
\right) \right] \; ,
\nonumber \\
\widetilde{\Delta}^{}_{32} \hspace{-0.15cm} & \simeq &
\hspace{-0.15cm} \Delta^{}_{31} \left[1 - \frac{1}{2} \alpha +
\frac{1}{2} \left(1 - 3|U^{}_{e 3}|^2 \right) \beta - \frac{1}{2}
\epsilon + \frac{3}{4} |U^{}_{e 3}|^2 \beta \left(2\beta +\epsilon
\right) \right] \; ;
\end{eqnarray}
and
\begin{eqnarray}
\Delta^{\prime}_{11} \hspace{-0.15cm} & \simeq & \hspace{-0.15cm}
-\Delta^{}_{31} \left[ \frac{1}{2} \alpha - \frac{1}{2} \left(1 -
|U^{}_{e 3}|^2\right) \beta - \frac{1}{2} \epsilon - \frac{1}{4}
|U^{}_{e 3}|^2 \beta \left(2\beta - 3\epsilon\right) \right] \; ,
\nonumber \\
\Delta^{\prime}_{12} \hspace{-0.15cm} & \simeq & \hspace{-0.15cm}
-\Delta^{}_{31} \left[ \frac{1}{2} \alpha - \frac{1}{2} \left(1 -
|U^{}_{e 3}|^2\right) \beta + \frac{1}{2} \epsilon - \frac{1}{4}
|U^{}_{e 3}|^2 \beta \left(2\beta + 3\epsilon\right) \right] \; ,
\nonumber \\
\Delta^\prime_{13} \hspace{-0.15cm} & \simeq & \hspace{-0.15cm}
-\Delta^{}_{31} \left(1 - |U^{}_{e 3}|^2 \beta + |U^{}_{e 3}|^2
\beta^2 \right) \; ,
\nonumber \\
\Delta^{\prime}_{21} \hspace{-0.15cm} & \simeq & \hspace{-0.15cm}
+\Delta^{}_{31} \left[ \frac{1}{2} \alpha + \frac{1}{2} \left(1 -
|U^{}_{e 3}|^2\right) \beta + \frac{1}{2} \epsilon + \frac{1}{4}
|U^{}_{e 3}|^2 \beta \left(2\beta - 3\epsilon\right) \right] \; ,
\nonumber \\
\Delta^{\prime}_{22} \hspace{-0.15cm} & \simeq & \hspace{-0.15cm}
+\Delta^{}_{31} \left[ \frac{1}{2} \alpha + \frac{1}{2} \left(1 -
|U^{}_{e 3}|^2\right) \beta - \frac{1}{2} \epsilon + \frac{1}{4}
|U^{}_{e 3}|^2 \beta \left(2\beta + 3\epsilon\right) \right] \; ,
\nonumber \\
\Delta^\prime_{23} \hspace{-0.15cm} & \simeq & \hspace{-0.15cm}
-\Delta^{}_{31} \left(1 - \alpha - |U^{}_{e 3}|^2 \beta + |U^{}_{e
3}|^2 \beta^2 \right) \; ,
\nonumber \\
\Delta^{\prime}_{31} \hspace{-0.15cm} & \simeq & \hspace{-0.15cm}
+\Delta^{}_{31} \left[ 1 - \frac{1}{2} \alpha + \frac{1}{2} \left(1
- |U^{}_{e 3}|^2\right) \beta + \frac{1}{2} \epsilon + \frac{1}{4}
|U^{}_{e 3}|^2 \beta \left(2\beta - 3\epsilon\right) \right] \; ,
\nonumber \\
\Delta^{\prime}_{32} \hspace{-0.15cm} & \simeq & \hspace{-0.15cm}
+\Delta^{}_{31} \left[ 1 - \frac{1}{2} \alpha + \frac{1}{2} \left(1
- |U^{}_{e 3}|^2\right) \beta - \frac{1}{2} \epsilon + \frac{1}{4}
|U^{}_{e 3}|^2 \beta \left(2\beta + 3\epsilon\right) \right] \; ,
\nonumber \\
\Delta^\prime_{33} \hspace{-0.15cm} & \simeq & \hspace{-0.15cm}
+\Delta^{}_{31} \left(|U^{}_{e 3}|^2 \beta + |U^{}_{e 3}|^2 \beta^2
\right) \; ,
\end{eqnarray}
where
\begin{eqnarray}
\epsilon \equiv \sqrt{\displaystyle \alpha^2 + 2 \left(|U^{}_{e
1}|^2 - |U^{}_{e 2}|^2 \right) \alpha\beta + \left(1 - 2|U^{}_{e
3}|^2 \right) \beta^2} \;\; .
\end{eqnarray}
Note that the smallness of $|U^{}_{e 3}|$ is already implied in
making the above approximations. With the help of Eqs. (3) and (4),
the expressions in Eq. (2) can be simplified to
\begin{eqnarray}
|\widetilde{U}^{}_{e 1}|^2 \hspace{-0.15cm} & \simeq &
\hspace{-0.15cm} \displaystyle +\frac{\alpha + \beta + \epsilon}{2
\epsilon} |U^{}_{e 1}|^2 - \frac{\alpha - \beta - \epsilon}{2
\epsilon} |U^{}_{e 2}|^2 \; ,
\nonumber \\
|\widetilde{U}^{}_{e 2}|^2 \hspace{-0.15cm} & \simeq &
\hspace{-0.15cm} \displaystyle -\frac{\alpha + \beta - \epsilon}{2
\epsilon} |U^{}_{e 1}|^2 + \frac{\alpha - \beta + \epsilon}{2
\epsilon} |U^{}_{e 2}|^2 \; ,
\nonumber \\
|\widetilde{U}^{}_{e 3}|^2 \hspace{-0.15cm} & \simeq &
\hspace{-0.15cm} \displaystyle |U^{}_{e 3}|^2 \;
\end{eqnarray}
in the leading-order approximation
\footnote{In the next-to-leading-order approximation one may obtain
the analytical result $|\widetilde{U}^{}_{e 3}|^2 \simeq \left(1 -
2\beta\right) |U^{}_{e 3}|^2$. Since $\beta$ is of ${\cal
O}(10^{-4})$ as estimated in Eq. (7), $|\widetilde{U}^{}_{e 3}|^2
\simeq |U^{}_{e 3}|^2$ is actually an excellent approximation.}.
Given $A \sim 1.52 \times 10^{-4} ~{\rm eV}^2 ~ Y^{}_e \left(\rho/
{\rm g}/{\rm cm}^3\right) \left(E/{\rm GeV}\right) \simeq 1.98
\times 10^{-4} ~{\rm eV}^2 \left(E/{\rm GeV}\right)$ for a realistic
oscillation experiment~\cite{Shrock}, where $Y^{}_e \simeq 0.5$ is
the electron fraction and $\rho \simeq 2.6 ~{\rm g}/{\rm cm}^3$ is
the typical matter density for an antineutrino trajectory through
the Earth's crust
\footnote{As for the JUNO or RENO-50 experiment, whose baseline
length is much shorter as compared with those accelerator-based
long-baseline neutrino oscillation experiments, it might be more
appropriate to take a somewhat smaller value of $\rho$. This issue
will be addressed later.},
we find that $\beta$ is much smaller than
$\alpha$ in magnitude:
\begin{eqnarray}
\alpha \hspace{-0.15cm} & \simeq & \hspace{-0.15cm} \displaystyle
3.12 \times 10^{-2} \times \frac{\Delta^{}_{21}}{7.5 \times 10^{-5}
~{\rm eV}^2} \times \frac{\pm 2.4 \times 10^{-3} ~{\rm
eV}^2}{\Delta^{}_{31}} \; ,
\nonumber \\
\beta \hspace{-0.15cm} & \simeq & \hspace{-0.15cm} \displaystyle
3.29 \times 10^{-4} \times \frac{E}{\rm 4 ~MeV} \times \frac{\pm 2.4
\times 10^{-3} ~{\rm eV}^2}{\Delta^{}_{31}} \; .
\end{eqnarray}
In this case one may simplify the expression of $\epsilon$ in Eq.
(5) as $\epsilon \simeq \alpha + \left(|U^{}_{e 1}|^2 - |U^{}_{e
2}|^2 \right) \beta$ plus much smaller terms. Note that Eqs. (3),
(4) and (6) are valid for a normal neutrino mass ordering. If an
inverted neutrino mass ordering (i.e., $\Delta^{}_{31} <0$) is taken
into account, the corresponding expressions can simply be obtained
from the above equations with a straightforward replacement
$\epsilon \to -\epsilon$.

In the standard parametrization of $U$~\cite{PDG}, $|U^{}_{e 1}| =
\cos\theta^{}_{12} \cos\theta^{}_{13}$, $|U^{}_{e 2}| =
\sin\theta^{}_{12} \cos\theta^{}_{13}$ and $|U^{}_{e 3}| =
\sin\theta^{}_{13}$. A global analysis of current neutrino
oscillation data yields the best-fit values $\theta^{}_{12} \simeq
33.5^\circ$ and $\theta^{}_{13} \simeq
8.5^\circ$~\cite{Fogli,Fogli2,Valle,GG}, which are insensitive to
the neutrino mass ordering. Therefore, $\epsilon \simeq \alpha +
\beta \cos 2\theta^{}_{12}$ holds as a good approximation. Taking
the same parametrization for the effective neutrino mixing matrix
$\widetilde{U}$ in matter, one may link the effective flavor mixing
angles $\widetilde{\theta}^{}_{12}$ and $\widetilde{\theta}^{}_{13}$
with the fundamental flavor mixing angles $\theta^{}_{12}$ and
$\theta^{}_{13}$ via Eq. (6). Namely,
\begin{eqnarray}
|\widetilde{U}^{}_{e 1}|^2 \hspace{-0.15cm} & \simeq &
\hspace{-0.15cm} \displaystyle \frac{\alpha + \beta
\cos^2\theta^{}_{12}}{\alpha + \beta \cos 2\theta^{}_{12}} |U^{}_{e
1}|^2 + \frac{\beta \cos^2\theta^{}_{12}}{\alpha + \beta \cos
2\theta^{}_{12}} |U^{}_{e 2}|^2 \; ,
\nonumber \\
|\widetilde{U}^{}_{e 2}|^2 \hspace{-0.15cm} & \simeq &
\hspace{-0.15cm} \displaystyle \frac{\alpha - \beta
\sin^2\theta^{}_{12}}{\alpha + \beta \cos 2\theta^{}_{12}} |U^{}_{e
2}|^2 - \frac{\beta \sin^2\theta^{}_{12}}{\alpha + \beta \cos
2\theta^{}_{12}} |U^{}_{e 1}|^2 \; ,
\nonumber \\
|\widetilde{U}^{}_{e 3}|^2 \hspace{-0.15cm} & \simeq &
\hspace{-0.15cm} \displaystyle |U^{}_{e 3}|^2 \; ;
\end{eqnarray}
and thus we arrive at the $\widetilde{\theta}^{}_{13} \simeq
\theta^{}_{13}$ and
\begin{eqnarray}
\cos^2\widetilde{\theta}^{}_{12} \hspace{-0.15cm} & \simeq &
\hspace{-0.15cm} \displaystyle \frac{\left(\alpha + \beta\right)
\cos^2\theta^{}_{12}}{\alpha + \beta \cos 2\theta^{}_{12}} \; ,
\nonumber \\
\sin^2\widetilde{\theta}^{}_{12} \hspace{-0.15cm} & \simeq &
\hspace{-0.15cm} \displaystyle \frac{\left(\alpha - \beta\right)
\sin^2\theta^{}_{12}}{\alpha + \beta \cos 2\theta^{}_{12}} \; .
\end{eqnarray}
Accordingly, we are left with
\begin{eqnarray}
\cos 2\widetilde{\theta}^{}_{12} \simeq \frac{\alpha \cos
2\theta^{}_{12} + \beta}{\alpha + \beta \cos 2\theta^{}_{12}} \simeq
\cos 2\theta^{}_{12} + \frac{A}{\Delta^{}_{21}} \sin^2\theta^{}_{12}
\; ,
\end{eqnarray}
and
\begin{eqnarray}
\sin^2 2\widetilde{\theta}^{}_{12} \simeq \frac{\left(\alpha^2 -
\beta^2\right) \sin^2 2\theta^{}_{12}}{\left(\alpha + \beta \cos
2\theta^{}_{12}\right)^2} \simeq \sin^2 2\theta^{}_{12} \left(1 -
2\frac{A}{\Delta^{}_{21}} \cos 2\theta^{}_{12}\right) \; ,
\end{eqnarray}
which are associated with a determination of the sign of
$\Delta^{}_{31}$ and with a precision measurement of the value of
$\theta^{}_{12}$, respectively. Note that Eqs.~(8)---(11) are valid
no matter whether the neutrino mass ordering is normal or inverted.
We see that the matter-induced correction is clearly characterized
by the ratio
\begin{eqnarray}
\frac{A}{\Delta^{}_{21}} \simeq 1.05 \times 10^{-2} \times
\frac{E}{\rm 4 ~MeV} \times \frac{7.5 \times 10^{-5} ~{\rm
eV}^2}{\Delta^{}_{21}} \; .
\end{eqnarray}
Therefore, we conclude that the precision measurements to be carried
out at JUNO and RENO-50 may suffer from the terrestrial matter
contamination at the $1\%$ level.

We proceed to calculate the matter-induced correction to the
probability of $\overline{\nu}^{}_e \to \overline{\nu}^{}_e$
oscillations. In vacuum, we have $P(\overline{\nu}^{}_e \to
\overline{\nu}^{}_e) = 1 - P^{}_{0} - P^{}_{*}$ with
\cite{WangXing}
\begin{eqnarray}
P^{}_0 \hspace{-0.15cm} & = & \hspace{-0.15cm} \displaystyle \sin^2
2 \theta^{}_{12} \cos^4 \theta^{}_{13} \sin^2 F^{}_{21} \nonumber
\\
P^{}_* \hspace{-0.15cm} & = & \hspace{-0.15cm} \displaystyle
\frac{1}{2} \sin^2 2 \theta^{}_{13} \left( 1 - \cos F^{}_* \cos
F^{}_{21} + \cos 2 \theta^{}_{12} \sin F^{}_* \sin F^{}_{21} \right)
\; ,
\end{eqnarray}
where $F^{}_{ji} \equiv 1267 \times \Delta^{}_{ji} L/E$ with
$\Delta^{}_{ji}$ being the neutrino mass-squared difference in unit
of $\rm eV^2$, $L$ being the baseline length in unit of km and $E$
being the antineutrino beam energy in unit of MeV (for $ji = 21, 31,
32$), and
\begin{eqnarray}
F^{}_* \equiv F^{}_{31} + F^{}_{32} = 1267 \times \frac{L}{E}
\left(\Delta^{}_{31} + \Delta^{}_{32}\right) = 1267 \times
\frac{L}{E} \Delta^{}_* \;
\end{eqnarray}
with the definition $\Delta^{}_* \equiv \Delta^{}_{31} +
\Delta^{}_{32}$. Needless to say, $\Delta^{}_*$ must be positive (or
negative) if the neutrino mass ordering is normal (or inverted).
Exactly parallel with Eq. (13), the expression of
$\widetilde{P}(\overline{\nu}^{}_e \to \overline{\nu}^{}_e)$ in
matter can be written as $\widetilde{P}(\overline{\nu}^{}_e \to
\overline{\nu}^{}_e) = 1 - \widetilde{P}^{}_0 - \widetilde{P}^{}_*$
with
\begin{eqnarray}
\widetilde{P}^{}_0 \hspace{-0.15cm} & = & \hspace{-0.15cm}
\displaystyle \sin^2 2 \widetilde{\theta}^{}_{12} \cos^4
\widetilde{\theta}^{}_{13} \sin^2 \widetilde{F}^{}_{21} \nonumber
\\
\widetilde{P}^{}_* \hspace{-0.15cm} & = & \hspace{-0.15cm}
\displaystyle \frac{1}{2} \sin^2 2\widetilde{\theta}^{}_{13} \left(
1 - \cos\widetilde{F}^{}_* \cos\widetilde{F}^{}_{21} + \cos
2\widetilde{\theta}^{}_{12} \sin\widetilde{F}^{}_*
\sin\widetilde{F}^{}_{21} \right) \; ,
\end{eqnarray}
where $\widetilde{F}^{}_{ji} \equiv 1267 \times
\widetilde{\Delta}^{}_{ji} L/E$ with $\widetilde{\Delta}^{}_{ji}$
being the effective neutrino mass-squared difference (for $ji = 21,
31, 32$), and
\begin{eqnarray}
\widetilde{F}^{}_* \equiv \widetilde{F}^{}_{31} +
\widetilde{F}^{}_{32} = 1267 \times \frac{L}{E}
\left(\widetilde{\Delta}^{}_{31} + \widetilde{\Delta}^{}_{32}\right)
= 1267 \times \frac{L}{E} \widetilde{\Delta}^{}_* \;
\end{eqnarray}
with the definition $\widetilde{\Delta}^{}_* \equiv
\widetilde{\Delta}^{}_{31} + \widetilde{\Delta}^{}_{32}$. With the
help of Eq. (3), we find that $\widetilde{\Delta}^{}_{21}$ and
$\widetilde{\Delta}^{}_*$ can approximate to
\begin{eqnarray}
\widetilde{\Delta}^{}_{21} \simeq
\Delta^{}_{21} + A \cos 2\theta^{}_{12} \; , \hspace{1cm}
\widetilde{\Delta}^{}_* \simeq \Delta^{}_* + A \; ,
\end{eqnarray}
respectively. Then Eq. (15) can be explicitly expressed as
\begin{eqnarray}
\widetilde{P}^{}_0 \hspace{-0.15cm} & \simeq & \hspace{-0.15cm}
\displaystyle P^{}_0 + A \sin^2 2 \theta^{}_{12} \cos
2\theta^{}_{12} \cos^4 \theta^{}_{13} \left(1267 \frac{L}{E} \sin 2
F^{}_{21} - \frac{2}{\Delta^{}_{21}} \sin^2 F^{}_{21} \right)
\nonumber \\
\widetilde{P}^{}_* \hspace{-0.15cm} & \simeq & \hspace{-0.15cm}
\displaystyle P^{}_* + \frac{1}{2} A \sin^2 2\theta^{}_{13} \left\{
1267\frac{L}{E} \left[\left(1 + \cos^2 2\theta^{}_{12}\right) \sin
F^{}_* \cos F^{}_{21} + 2 \cos 2\theta^{}_{12} \cos F^{}_* \sin
F^{}_{21} \right] \right.
\nonumber \\
& & \hspace{-0.15cm} \displaystyle \left. \hspace{3.45cm}
+ \frac{1}{\Delta^{}_{21}}
\sin^2 \theta^{}_{12} \sin F^{}_* \sin F^{}_{21} \right\} \; ,
\end{eqnarray}
where $F^{}_{21} = 1267 \Delta^{}_{21} L/E \sim \pi/2$ (or
equivalently, $L \sim 50$ km) has been implied in accordance with
the designs of the JUNO~\cite{JUNO,JUNOCDR} and RENO-50~\cite{RENO}
experiments, and hence $1267 A L/E \sim A/\Delta^{}_{21} \sim
10^{-2}$ is a small expansion parameter. The difference
\begin{eqnarray}
\widetilde{P}(\overline{\nu}^{}_e \to \overline{\nu}^{}_e) -
P(\overline{\nu}^{}_e \to \overline{\nu}^{}_e) = \left(P^{}_0 -
\widetilde{P}^{}_0 \right) + \left(P^{}_* -
\widetilde{P}^{}_*\right) \; ,
\end{eqnarray}
which is proportional to $A$ as shown in Eq. (18), is therefore a
clear measure of the terrestrial matter effects associated with JUNO
or RENO-50.

\vspace{0.4cm}

Now we turn to a numerical study of the terrestrial matter effects
in a medium-baseline reactor antineutrino oscillation experiment
like JUNO or RENO-50. For simplicity and illustration, we adopt the
best-fit values $\Delta^{}_{21} \simeq 7.5 \times 10^{-5}~{\rm
eV}^2$, $\Delta^{}_{*} \simeq 4.839 \times 10^{-3}~{\rm eV}^2$,
$\sin^2\theta^{}_{12} \simeq 0.304$ and $\sin^2\theta^{}_{13} \simeq
0.0218$ obtained from a recent global analysis of current neutrino
oscillation data~\cite{GG}. The terrestrial matter density along the
antineutrino trajectory is typically assumed to be $\rho \simeq 2.6
~{\rm g/cm}^3$, and its uncertainty will be briefly discussed later
on. In our analysis we are going to focus on the {\it normal}
neutrino mass ordering as the {\it true} mass ordering, and we find
that our main conclusion will actually keep valid even if the
inverted neutrino mass ordering is taken into account.
\begin{figure}
\begin{center}
\begin{tabular}{cc}
\includegraphics*[bb=50 30 720 520, width=0.47\textwidth]{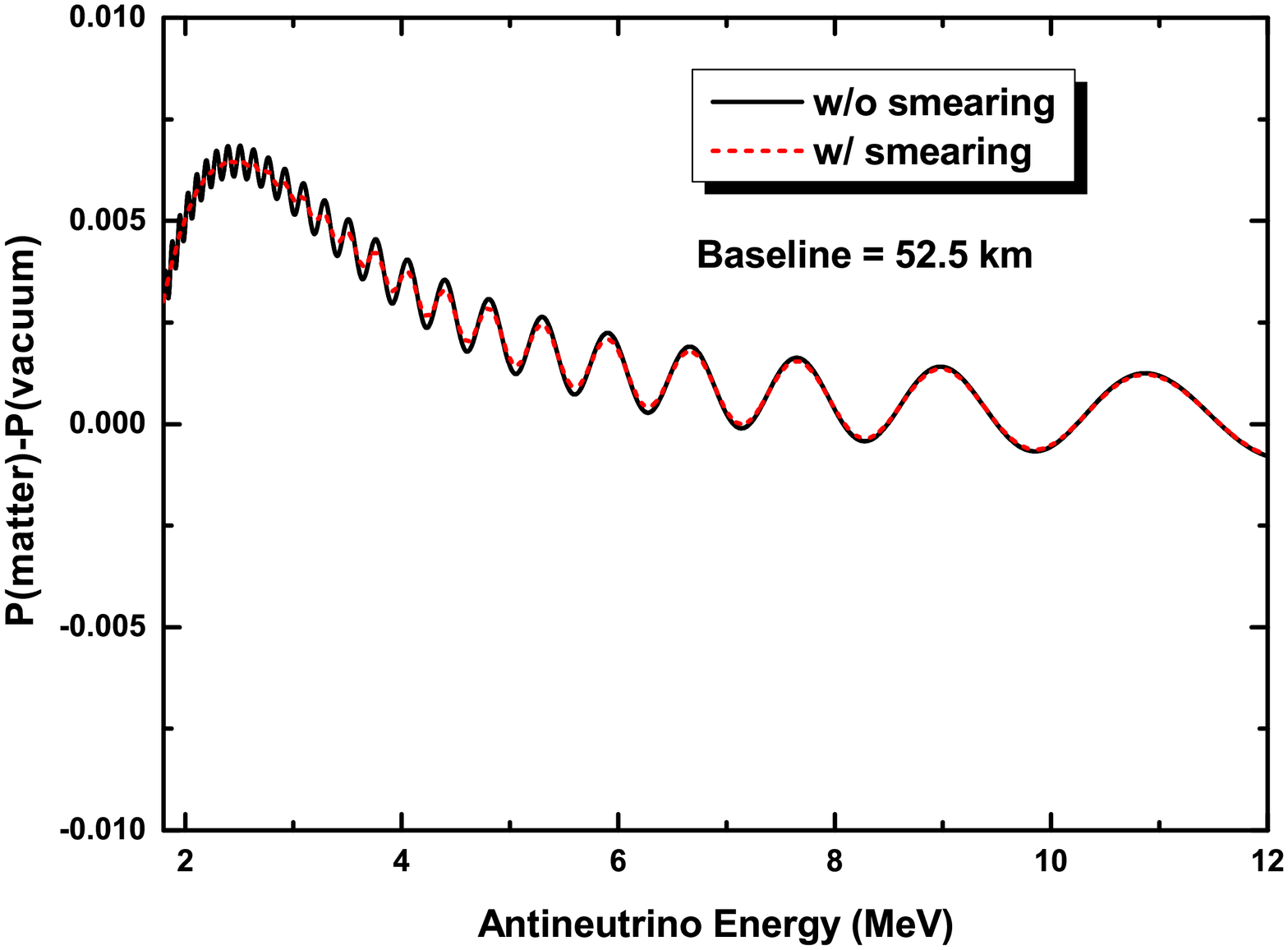}
&
\includegraphics*[bb=50 30 720 520, width=0.47\textwidth]{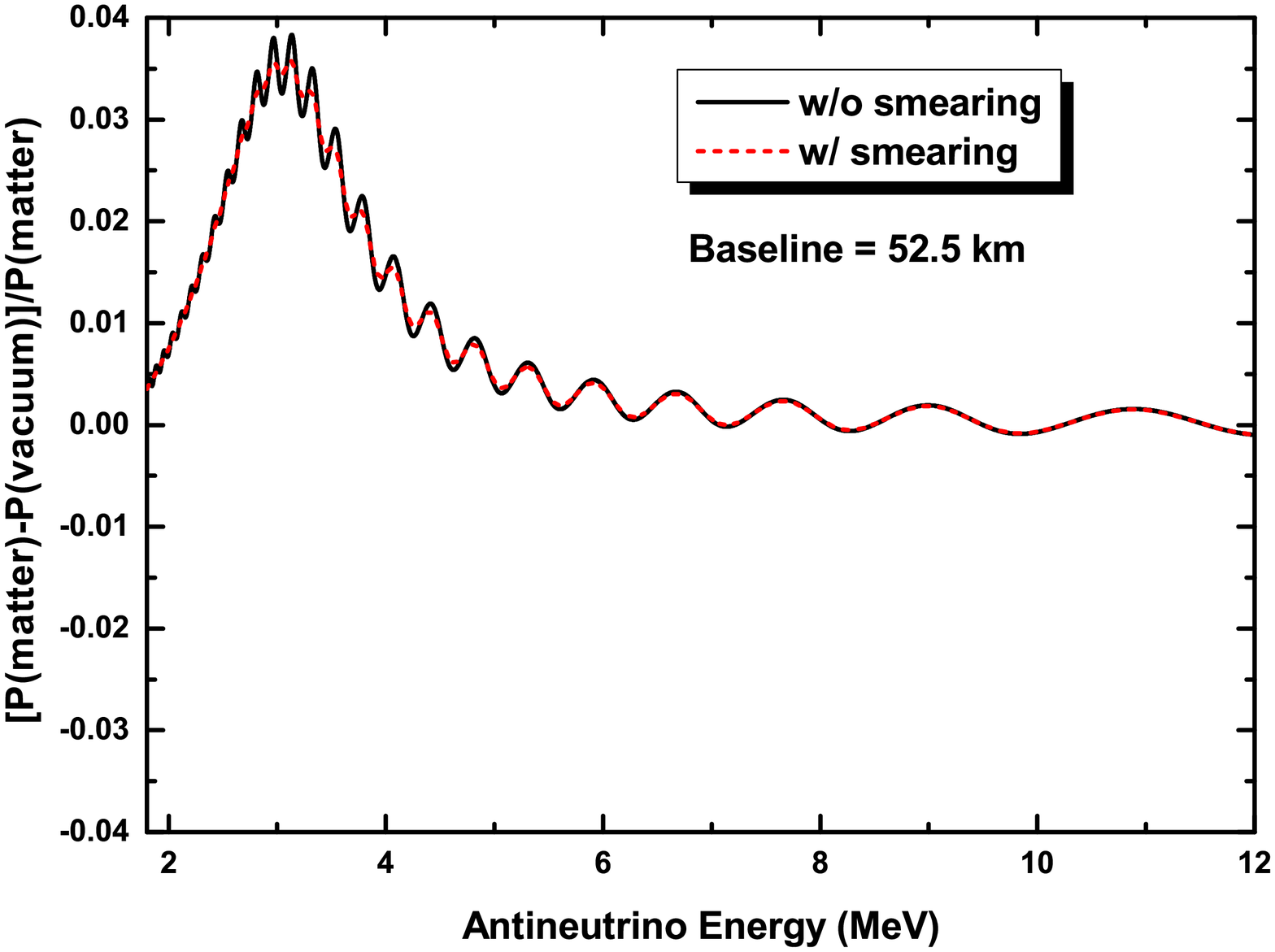}
\end{tabular}
\end{center}
\vspace{-0.6cm} \caption{The absolute (left panel) and relative
(right panel) differences between $\widetilde{P}(\overline{\nu}^{}_e
\to \overline{\nu}^{}_e)$ (in matter) and $P(\overline{\nu}^{}_e \to
\overline{\nu}^{}_e)$ (in vacuum) for a reactor antineutrino
oscillation experiment with $L = 52.5$ km. The solid lines
correspond to the true antineutrino energy, and the dashed lines are
averaged over a Gaussian energy resolution of 3\%/$\sqrt{E\,(\rm
MeV)}$. \label{fig:diff1}}
\end{figure}

As a result of our exact numerical calculations without involving
any analytical approximations, Fig.~\ref{fig:diff1} shows the
absolute (left panel) and relative (right panel) differences between
the matter-corrected probability $\widetilde{P}(\overline{\nu}^{}_e
\to \overline{\nu}^{}_e)$ and its vacuum counterpart
$P(\overline{\nu}^{}_e \to \overline{\nu}^{}_e)$ associated with a
medium-baseline ($L = 52.5$ km) reactor antineutrino oscillation
experiment. The solid curves are for the true antineutrino energy,
and the dashed ones are averaged over a Gaussian energy resolution
of 3\%/$\sqrt{E\,(\rm MeV)}$. We see that the absolute
difference $\widetilde{P}(\overline{\nu}^{}_e \to
\overline{\nu}^{}_e) - P(\overline{\nu}^{}_e \to
\overline{\nu}^{}_e)$ can reach about $0.7\%$ in the vicinity of the
first oscillation peak of $\Delta^{}_{21}$, which corresponds to a
relative matter-induced correction of about $4\%$ illustrated on the
right panel of Fig. 1. As a matter of fact, the main profile of
$\widetilde{P}(\overline{\nu}^{}_e \to \overline{\nu}^{}_e) -
P(\overline{\nu}^{}_e \to \overline{\nu}^{}_e)$ or
$\left[\widetilde{P}(\overline{\nu}^{}_e \to \overline{\nu}^{}_e) -
P(\overline{\nu}^{}_e \to \overline{\nu}^{}_e)\right]/
\widetilde{P}(\overline{\nu}^{}_e \to \overline{\nu}^{}_e)$ is
attributed to the $\Delta^{}_{21}$-triggered oscillation, where the
matter-induced suppression in $\sin^2 2\widetilde{\theta}^{}_{12}$
provides a positive correction in the $\Delta^{}_{21}$-dominated
range. The small wiggles in Fig. 1 are caused by the
$\Delta^{}_{*}$-triggered oscillation, and their amplitudes are
modulated by the energy-dependent correction of $\cos
2\widetilde{\theta}^{}_{12}$.
\begin{figure}
\begin{center}
\begin{tabular}{cc}
\includegraphics*[bb=50 30 720 520, width=0.47\textwidth]{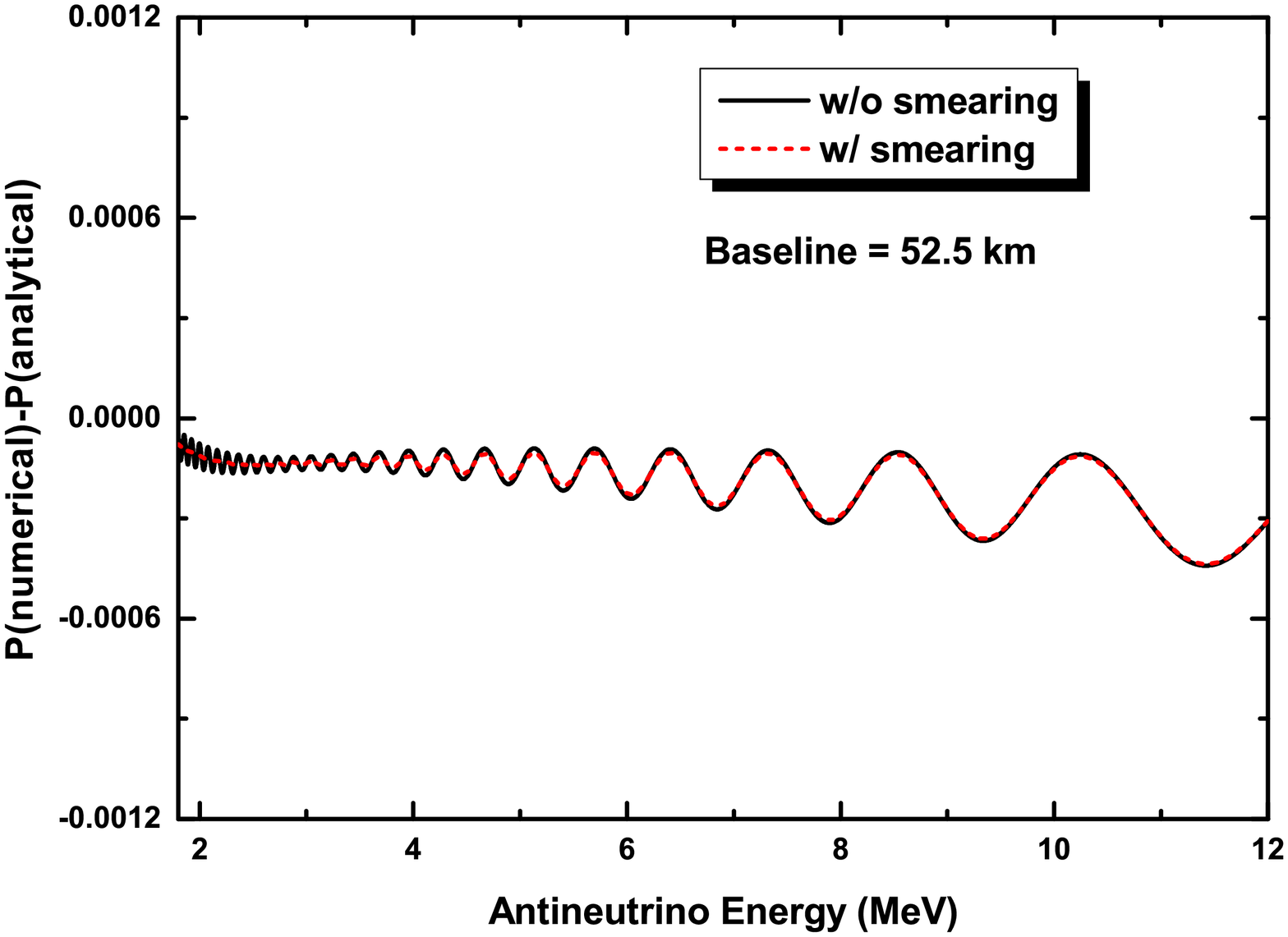}
&
\includegraphics*[bb=50 30 720 520, width=0.47\textwidth]{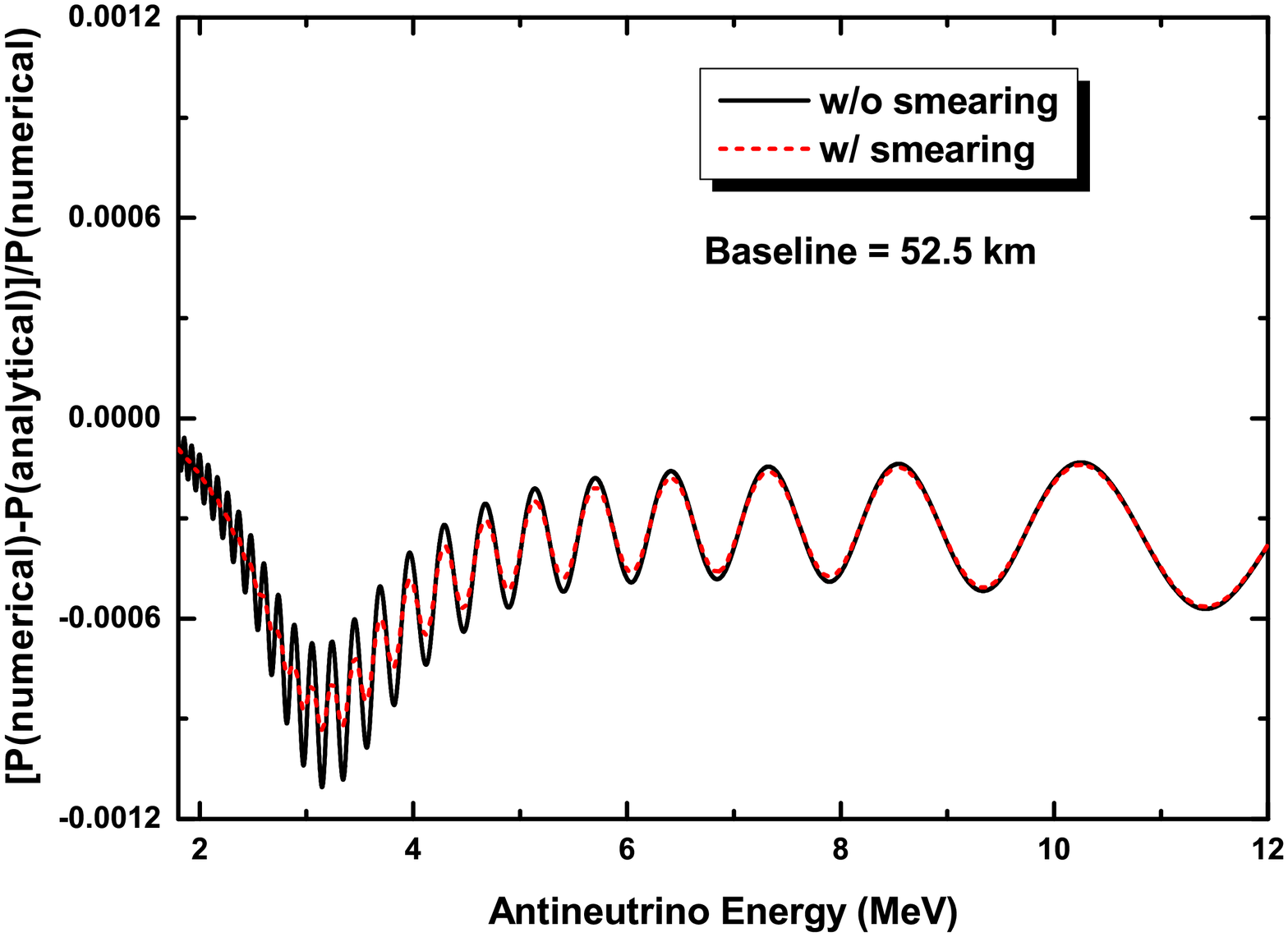}
\end{tabular}
\end{center}
\vspace{-0.6cm} \caption{A comparison between the results of
$\widetilde{P}(\overline{\nu}^{}_e \to \overline{\nu}^{}_e)$
achieved from an exact numerical calculation (numerical) and from
the analytical approximations in Eq. (18) (analytical): their
absolute (left panel) and relative (right panel) differences with or
without smearing for a reactor antineutrino oscillation experiment
with $L = 52.5$ km. The solid curves are for the true antineutrino
energy, and the dashed ones are averaged over a Gaussian energy
resolution of 3\%/$\sqrt{E\,(\rm MeV)}$. \label{fig:diff2}}
\end{figure}

Before calculating the statistical sensitivity of a realistic
experimental measurement, it is necessary to test the accuracy of
our analytical approximations made in Eqs.~(8)---(11) and
Eqs.~(15)---(18). Fig.~\ref{fig:diff2} shows a comparison between
the results of $\widetilde{P}(\overline{\nu}^{}_e \to
\overline{\nu}^{}_e)$ obtained from a complete numerical calculation
and the analytical approximations made in Eq. (18): their absolute
(left panel) and relative (right panel) differences with or without
smearing effects for a reactor antineutrino oscillation experiment
like JUNO or RENO-50. In this figure the solid lines are for the
true antineutrino energy, and the dashed curves are averaged over a
Gaussian energy resolution of 3\%/$\sqrt{E\,(\rm MeV)}$. We find
that the absolute errors of our analytical approximations are lower
than $3\times 10^{-4}$ in most of the antineutrino energy range,
proving that Eq.~(18) and the associated analytical approximations
can be safely employed in the following sensitivity studies.

Taking account of JUNO's nominal setup as described in
Refs.~\cite{JUNO,Li:2013zyd}, we are going to illustrate how the
terrestrial matter effects influence the measurements of both the
neutrino mass ordering and the flavor mixing parameters. We shall
also discuss an important issue: to what extent one can establish or
constrain the terrestrial matter effects at JUNO or RENO-50, or in a
similar experiment to be proposed.

Given the JUNO simulation, which has been described in detail in
Ref. \cite{JUNO}, let us consider a 20 kt liquid scintillator
detector with the energy resolution of 3\%/$\sqrt{E\,(\rm MeV)}$
\footnote{A generic parametrization of the energy resolution is
written as $\sqrt{({a}/{\sqrt{E}})^2+b^2+({c}/{E})^2}$, which is
numerically equivalent to an effective energy resolution of
$\sqrt{{a}^2+(1.6\times b)^2+(c/1.6)^2}/{\sqrt{E}}$ in the mass
ordering measurement~\cite{JUNO}. The requirement of
3\%/$\sqrt{E\,(\rm MeV)}$ can be regarded as the total contribution
of all the stochastic and non-stochastic terms. }.
We take account of the real reactor powers and baseline
distributions of the Yangjiang and Taishan nuclear power plants
listed in Table~2 of Ref.~\cite{JUNO}, which have a total thermal
power of 36 ${\rm GW}_{\rm th}$ and a power-weighted baseline of
52.5 km. Moreover, we assume a detection efficiency of 80\% and the
nominal running time of six years and 300 effective days per year in
our numerical simulation.

To discuss the statistical sensitivity of the experimental
measurement
\footnote{See
Refs.~\cite{JUNO,Li:2013zyd,MHsens1,MHsens2,Ciuffoli:2013ep,MHsens3,RENO50sens}
for an incomplete list of the works dealing with the statistical
sensitivity of the mass ordering measurement in a medium-baseline
reactor antineutrino experiment.},
we construct the following standard $\chi^2$ function:
\begin{equation}
\chi^2_{} = 
\sum^{N^{}_{\text{bin}}}_{i=1}
\frac{\left[M^{}_{i}(p^{M},\eta) - T^{}_{i}(p^{T},\eta)\left(1+
\displaystyle\sum_k
\alpha^{}_{ik}\epsilon^{}_{k}\right)\right]^2}{M^{}_{i}(p^{M},\eta)}
+ \sum_k\frac{\epsilon^2_{k}}{\sigma^2_k} \, ,
\label{chiREA}
\end{equation}
where $M^{}_{i}$ and $T^{}_{i}$ are the measured and predicted
antineutrino events in the $i$-th antineutrino energy bin,
respectively; $\sigma_k$ and $\epsilon_{k}$ are the $k$-th
systematic uncertainty and the corresponding pull parameter,
respectively. The considered nominal systematic uncertainties
include the correlated reactor rate uncertainty ($\sim 2\%$), the
uncorrelated reactor rate uncertainty ($\sim 0.8\%$), the
energy-uncorrelated bin-to-bin reactor flux spectrum uncertainty
($\sim 1\%$) and the detector-related uncertainty ($\sim 1\%$). Some
additional important systematic uncertainties on the measurements of
the neutrino mass ordering and oscillation parameters have been
thoroughly discussed in sections 2 and 3 of Ref.~\cite{JUNO}. In
Eq.~(\ref{chiREA}), $p$ stands for the oscillation parameters (i.e.,
$p= \left\{\Delta^{}_{21}, \Delta^{}_{*}, \theta^{}_{12},
\theta^{}_{13} \right\}$), and $\eta \equiv A(\rho)/A(\rho=2.6\;{\rm
g/cm}^3)$ is defined as the effective matter potential index.
\begin{figure}
\begin{center}
\begin{tabular}{c}
\includegraphics*[bb=80 30 730 520,width=0.6\textwidth]{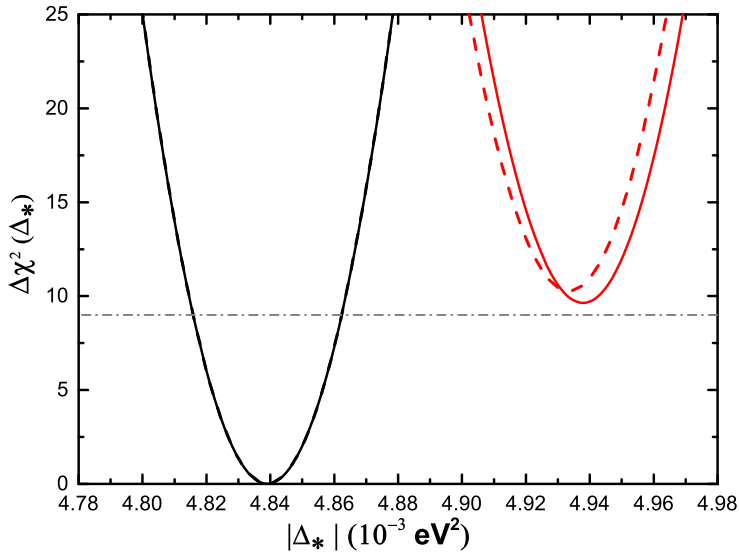}
\end{tabular}
\end{center}
\vspace{-0.6cm} \caption{A comparison of the neutrino mass ordering
sensitivities with (red solid) and without (red dashed) considering
the terrestrial matter effects. The vertical distance (defined as
$\Delta\chi^2_{\rm MO}$) between the minima of the red and black
lines denotes the sensitivity of the mass ordering measurement.
\label{fig:mh1}}
\end{figure}

\begin{figure}
\begin{center}
\begin{tabular}{c}
\includegraphics*[bb=60 40 720 520,width=0.6\textwidth]{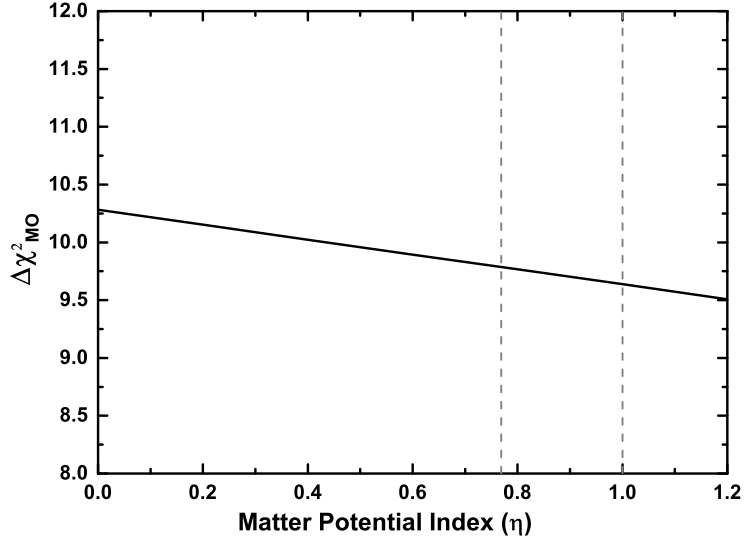}
\end{tabular}
\end{center}
\vspace{-0.6cm} \caption{An illustration of the sensitivity of the
mass ordering measurement $\Delta\chi^2_{\rm MO}$ as a function of
the matter potential index $\eta \equiv A(\rho)/A(\rho=2.6\;{\rm
g/cm}^3)$. The vertical dashed line with $\eta\simeq1$ or $0.77$
stands for the terrestrial matter density $\rho \simeq 2.6\;{\rm
g/cm}^3$ or $2.0\;{\rm g/cm}^3$, respectively. The value of
$\Delta\chi^2_{\rm MO}$ for $\eta\simeq 0$, $0.77$ or $1$ is
$10.28$, $9.79$ or $9.64$, respectively. \label{fig:mh2}}
\end{figure}

Fig.~\ref{fig:mh1} is a comparison of the neutrino mass ordering
sensitivities with (solid) and without (dashed) considering the
terrestrial matter effects. The black lines come from the fitting in
the assumption of the normal mass ordering (NMO) of three neutrinos,
and the red lines assume the inverted mass ordering (IMO). The
vertical distance between the minima of the red and black curves is
defined as a measure of the neutrino mass ordering sensitivity:
\begin{equation}
\Delta \chi^2_{\text{MO}} = \left|\chi^2_{\rm min}(\rm NMO) -
\chi^2_{\rm min}(\rm IMO) \right| \; ,
\end{equation}
where the minimization is implemented for all the relevant
oscillation and pull parameters. Compared with the situation of
$\overline{\nu}^{}_e \to \overline{\nu}^{}_e$ oscillations in
vacuum, the inclusion of terrestrial matter effects may reduce the
value of $\Delta\chi^2_{\rm MO}$ from 10.28 to 9.64, which is
comparable with other important systematic uncertainties and hence
should not be neglected in the future mass ordering measurement. In
the above calculation we have typically taken $\rho \simeq 2.6\;{\rm
g/cm}^3$ for the terrestrial matter density. For the reactor
antineutrino oscillations with a medium baseline (i.e., $L\sim 50$
km from the reactors to the detector), however, the
$\overline{\nu}^{}_e$ trajectory during propagation is expected to
include a large proportion of the sedimentary layer. In other words,
the realistic experiment may actually involve a somewhat smaller
terrestrial matter density. In Fig.~\ref{fig:mh2} we illustrate the
sensitivity of the mass ordering measurement $\Delta\chi^2_{\rm MO}$
as a function of the matter potential index $\eta$. One can see that
$\Delta\chi^2_{\rm MO}$ depends linearly on $\eta$. If a smaller
matter density $\rho \simeq 2.0\;{\rm g/cm}^3$ is taken into account
for JUNO, the mass ordering sensitivity reduction will be from 10.28
to 9.79.
\begin{figure}
\begin{center}
\begin{tabular}{cc}
\includegraphics*[bb=30 30 620 520,width=0.47\textwidth]{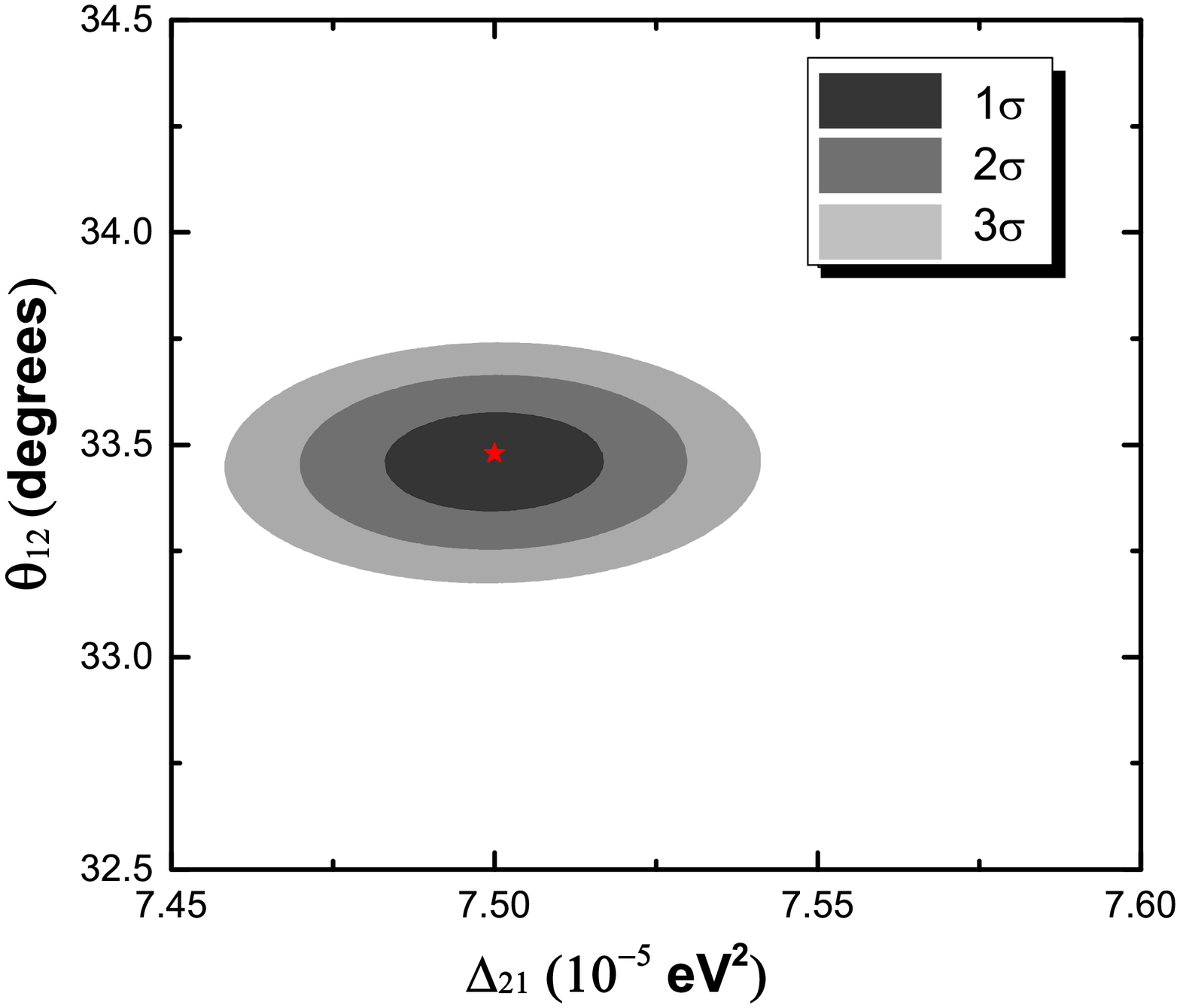}
&
\includegraphics*[bb=25 30 620 520,width=0.47\textwidth]{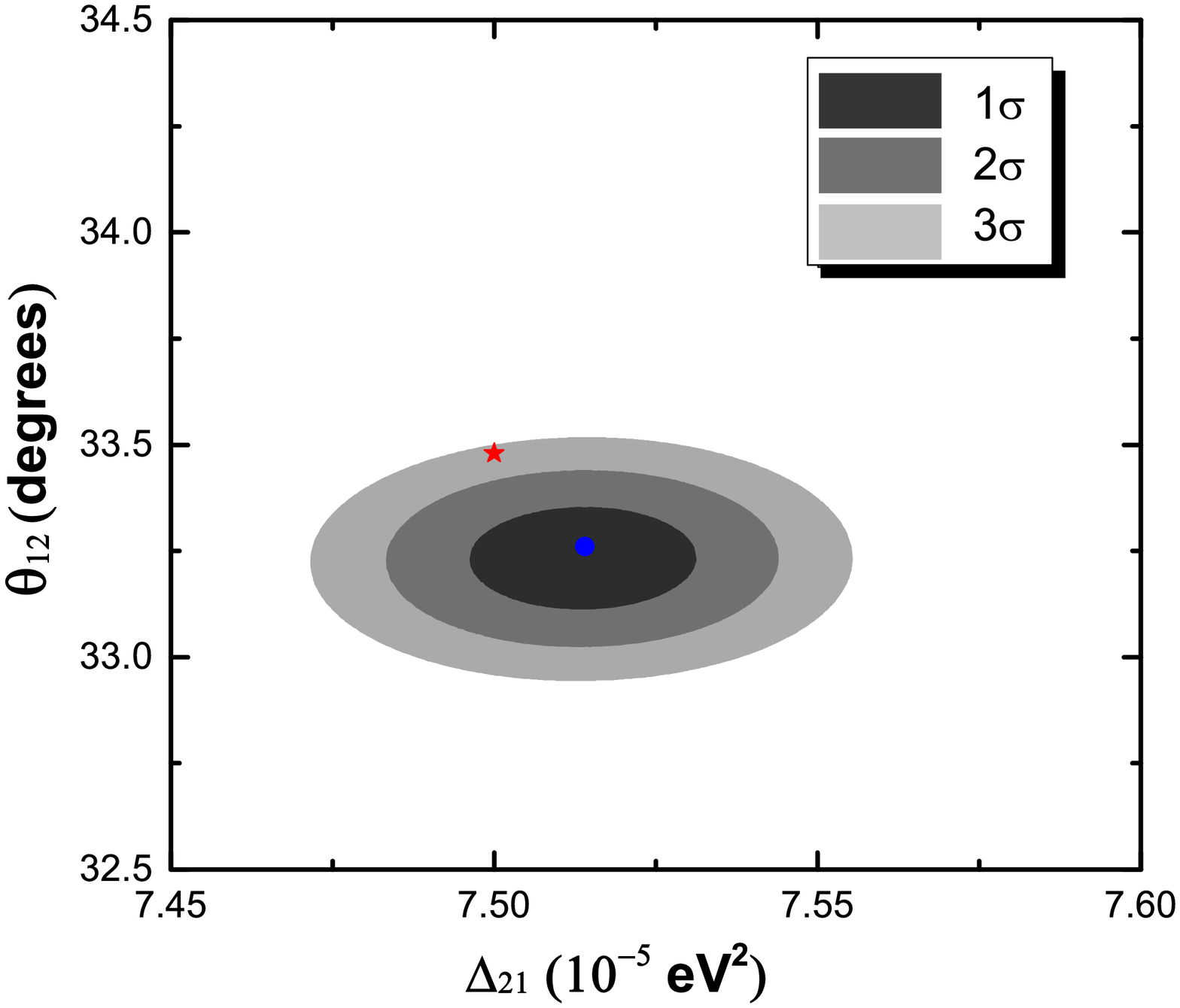}
\end{tabular}
\end{center}
\vspace{-0.6cm} \caption{The allowed regions of $\Delta^{}_{21}$ and
$\theta^{}_{12}$ with (left panel) and without (right panel)
including terrestrial matter effects in the predictions. The matter
density $\rho \simeq 2.6\;{\rm g/cm}^3$ is assumed in the
measurements. The red stars denote the true values of
$\Delta^{}_{21}$ and $\theta^{}_{12}$, and the blue dot is the
best-fit point when the terrestrial matter effects are omitted.
\label{fig:shift}}
\end{figure}

Now we turn to discuss the terrestrial matter effects on the
relevant flavor parameters. In our numerical analysis, $\rho \simeq
2.6\;{\rm g/cm}^3$ (i.e., $\eta \simeq 1$) is typically taken to
modulate the measured antineutrino events $M^{}_{i}$. In the left
panel of Fig.~\ref{fig:shift} we include terrestrial matter effects
in the predicted antineutrino events $T^{}_{i}$ and display the
fitting results of $\Delta^{}_{21}$ and $\theta^{}_{12}$. The red
star denotes the true values of these two parameters. It turns out
that the best-fit points can return to the true values, and the
allowed regions are consistent with the fitting results in the
assumption of the vacuum $\overline{\nu}^{}_e \to
\overline{\nu}^{}_e$ oscillations (see section~3.2 of
Ref.~\cite{JUNO}). The 1$\sigma$ precision levels of
$\Delta^{}_{21}$ and $\sin^2\theta^{}_{12}$ with the nominal
systematic setup can reach $0.23\%$ and $0.58\%$, respectively. In
comparison, the 1$\sigma$ precision levels of $\Delta^{}_{21}$ and
$\sin^2\theta^{}_{12}$ in the absence of matter effects were found
to be $0.24\%$ and $0.54\%$, respectively (see section~3.2 of
Ref.~\cite{JUNO}). A minor reduction in the accuracy of
$\sin^2\theta^{}_{12}$ is certainly attributed to the suppression of
$\theta^{}_{12}$ in terrestrial matter.

For the sake of comparison, let us neglect terrestrial matter
effects in the predicted antineutrino events $T_{i}$ and illustrate
the fitting results of $\Delta^{}_{21}$ and $\theta^{}_{12}$ in the
right panel of Fig.~\ref{fig:shift}. The red star points to the true
values of these two parameters, and the blue dot stands for the
best-fit point. The allowed regions are shifted to higher
$\Delta^{}_{21}$ and lower $\theta^{}_{12}$, and the best-fit point
is located at $\Delta^{}_{21} \simeq 7.514 \times 10^{-5}\; {\rm
eV}^2$ and $\theta^{}_{12} \simeq 33.26^\circ$. The precision of
$\Delta^{}_{21}$ and $\theta^{}_{12}$ turns out to be the same as
that in the left panel of Fig.~\ref{fig:shift}. Hence the best-fit
values of $\Delta^{}_{21}$ and $\theta^{}_{12}$ deviate around
0.8$\sigma$ and 2.4$\sigma$ from their true values, respectively. If
additional systematic uncertainties~\cite{JUNO} of the flux spectrum
and the energy scale are taken into account in the analysis, the
sizes of deviation might be more or less reduced.
\begin{figure}
\begin{center}
\begin{tabular}{c}
\includegraphics*[bb=80 30 720 520,width=0.65\textwidth]{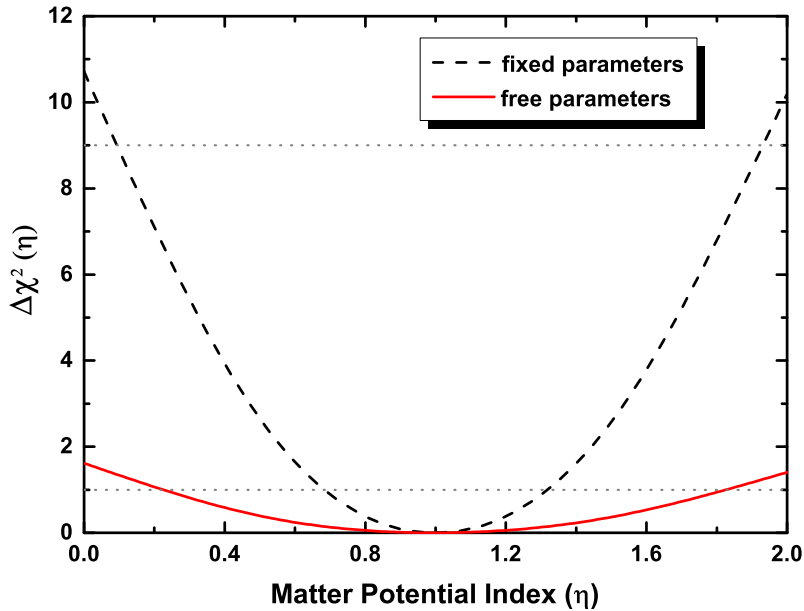}
\end{tabular}
\end{center}
\vspace{-0.6cm} \caption{An illustration of the sensitivity of the
terrestrial matter effects with the JUNO nominal setup. The black
dashed and red solid lines are shown for the fitting results without
and with considering the uncertainties of the neutrino oscillation
parameters, respectively. \label{fig:sens}}
\end{figure}

Finally let us discuss to what extent one can establish or constrain
the terrestrial matter effects at JUNO. Assuming a matter density
$\rho \simeq 2.6\;{\rm g/cm}^3$ in the measured antineutrino events,
we illustrate the change of $\Delta \chi^2(\eta)$ as a function of
the matter potential index $\eta$ in Fig.~\ref{fig:sens} with both
fixed and free oscillation parameters. If all the oscillation
parameters are fixed, we obtain $\Delta \chi^2(0)\simeq 11$,
indicating that the terrestrial matter effects can be tested with a
significance of more than 3$\sigma$. However, the significance of
establishing the terrestrial matter effects will significantly
reduce to 1.3$\sigma$ after the oscillation parameters are
marginalized. This can be understood with the help of Eqs.~(11) and
(17), where the corrections of the matter potential to
$\sin^2\theta^{}_{12}$ and $\Delta^{}_{21}$ are about $0.8\%$ and
$0.4\%$, respectively. If some additional systematic uncertainties
are considered in the analysis~\cite{JUNO}, including the
background, the reactor flux spectrum uncertainty of 1\%, the energy
scale uncertainty of 1\% and the energy non-linear uncertainty of
1\%, then the projected precision levels for $\sin^2\theta^{}_{12}$
and $\Delta^{}_{21}$ will be 0.72\% and 0.60\%, respectively.
Correspondingly, the sensitivity of establishing the terrestrial
matter effects will be less than 1$\sigma$.
\begin{figure}
\begin{center}
\begin{tabular}{c}
\includegraphics*[bb=80 30 720 520,width=0.64\textwidth]{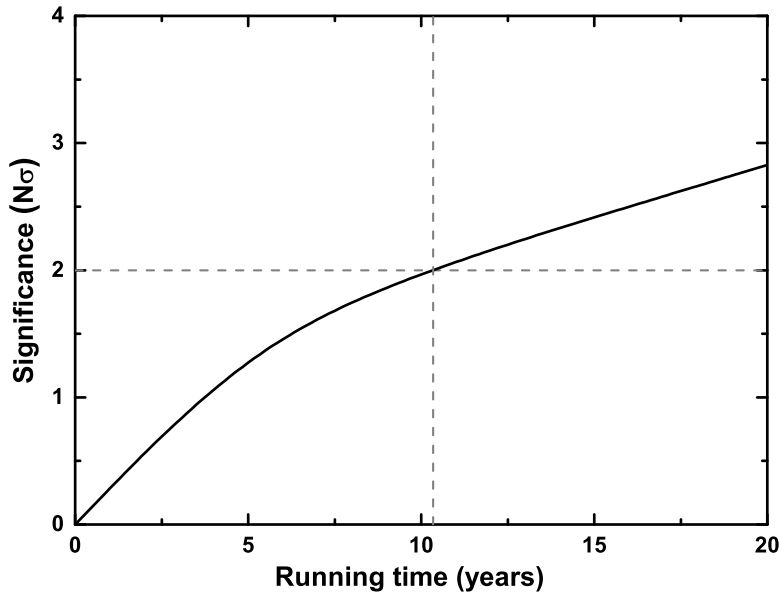}
\end{tabular}
\end{center}
\vspace{-0.6cm} \caption{An illustration of the sensitivity of
ruling out the vacuum neutrino oscillation scenario (i.e., $\eta=0$)
as a function of the running time with the nominal JUNO
configuration and appropriate near detectors. The significance is
defined as the squared root of ${\Delta \chi^2(\eta=0)}$.
\label{fig:2det}}
\end{figure}

If the near detectors can be built to monitor the reactor
antineutrino flux, a relative measurement of the rate and spectrum
between the near and far detectors is expected to significantly
reduce the reactor- and detector-related systematic uncertainties in
the $\sin^2\theta^{}_{12}$ and $\Delta^{}_{21}$ measurements, and
thus the sensitivity of establishing the terrestrial matter effects
can accordingly increase. Without specifying the details of near
detectors, we just split the systematic uncertainties into the
(detector-correlated) absolute uncertainties and
(detector-uncorrelated) relative uncertainties. Assuming the
absolute errors will be cancelled by virtue of near detectors and
the relative errors are at the Daya Bay
level~\cite{DYB1,DYB2,DYB3,DYB4}, we show the sensitivity of ruling
out the vacuum neutrino oscillation scenario (i.e., $\eta=0$) as a
function of the running time in Fig.~\ref{fig:2det}, where the
significance is defined as the squared root of ${\Delta
\chi^2(\eta=0)}$. We observe that a 2$\sigma$ sensitivity of
establishing the terrestrial matter effects can be achieved for
about 10 years of data taking, if one or two appropriate near
detectors are implemented to the nominal JUNO configuration. Further
details on the near detector configuration will be discussed
elsewhere
\footnote{Given different motivation and different detector
consideration, there are a few other works on the near detector
ideas for a medium-baseline reactor antineutrino oscillation
experiment~\cite{ND1,ND2,ND3}.}.

\vspace{0.4cm}

To summarize, we have examined how small the terrestrial matter
effects can be in a medium-baseline reactor antineutrino oscillation
experiment like JUNO or RENO-50, which aims to carry out a precision
measurement of the neutrino mass ordering and relevant flavor
parameters. To do so, we have expanded the probability of
$\overline{\nu}^{}_e \to \overline{\nu}^{}_e$ oscillations with $L
\simeq 50$ km in terms of the small matter parameter. Our analytical
approximations are simple but accurate enough for a deeper
understanding of the outputs of the exact numerical calculations.
Taking the ongoing JUNO experiment as a good example, we have shown
that the inclusion of terrestrial matter effects is likely to reduce
the sensitivity of the neutrino mass ordering measurement by $\Delta
\chi^2_{\rm MO}\simeq 0.6$. We find that the terrestrial matter
effects may also shift the best-fit values of $\theta^{}_{12}$ and
$\Delta^{}_{21}$ by about $1\sigma$ to $2\sigma$ if they are ignored
in the future data analysis.

We conclude that the terrestrial matter effects must be carefully
taken into account because they are non-negligible in the
reactor-based measurements of the neutrino mass ordering and
$\overline{\nu}^{}_e \to \overline{\nu}^{}_e$ oscillation
parameters. But it remains difficult to establish the profile of
terrestrial matter effects at a high significance level in a
realistic experiment of this kind, such as JUNO or RENO-50. This
issue motivates us to consider the possibility of installing the
near detectors to measure the initial reactor antineutrino flux
\footnote{We thank the useful communication with Jun Cao in this
connection.},
where the matter effects have not been developed. In this case a
comparison between the measurement of
$\widetilde{P}(\overline{\nu}^{}_e \to \overline{\nu}^{}_e)$ and its
energy dependence at the far detector ($L \simeq 50$ km) and that of
$P(\overline{\nu}^{}_e \to \overline{\nu}^{}_e)$ at the near
detectors ($L \sim 0$) will allow one to probe the fine effects of
terrestrial matter associated with JUNO or RENO-50. Our preliminary
estimate indicates that it is possible to establish the terrestrial
matter effects with a $2\sigma$ sensitivity for about 10 years of
data taking at JUNO with the help of a proper near detector
implementation.

\vspace{0.5cm}

We thank Eligio Lisi, Shun Zhou and Jing-yu Zhu for useful
discussions. This work was supported in part by the National Natural
Science Foundation of China under Grant Nos. 11135009 and 11305193,
by the Strategic Priority Research Program of the Chinese Academy of
Sciences under Grant No.~XDA10010100, and by the CAS Center for
Excellence in Particle Physics.



\begin{thebibliography}{99}
\bibitem{JUNO}
F.~An {\it et al.} [JUNO Collaboration],
Neutrino Physics with JUNO, J.\ Phys.\ G {\bf 43} (2016) 030401
[arXiv:1507.05613 [physics.ins-det]].

\bibitem{JUNOCDR}
Z.~Djurcic {\it et al.} [JUNO Collaboration], JUNO Conceptual Design
Report, (2015) arXiv:1508.07166 [physics.ins-det].

\bibitem{MHrev1}
X.~Qian and P.~Vogel, Neutrino Mass Hierarchy,
Prog.\ Part.\ Nucl.\ Phys.\  {\bf 83} (2015) 1  [arXiv:1505.01891 [hep-ex]].

\bibitem{MHrev2}
R.~B.~Patterson, Prospects for Measurement of the Neutrino Mass
Hierarchy, Ann.\ Rev.\ Nucl.\ Part.\ Sci.\  {\bf 65} (2015) 177
[arXiv:1506.07917 [hep-ex]].

\bibitem{RENO} S.~B.~Kim, New results from RENO and prospects
with RENO-50, Nucl.\ Part.\ Phys.\ Proc.\  {\bf 265}--{\bf 266} (2015) 93
[arXiv:1412.2199 [hep-ex]].

\bibitem{Lisi1} F.~Capozzi, E.~Lisi and A.~Marrone,
Neutrino mass hierarchy and electron neutrino oscillation parameters
with one hundred thousand reactor events, Phys.\ Rev.\ D {\bf 89}
(2014) 013001 [arXiv:1309.1638 [hep-ph]].

\bibitem{Li-JUNO} Y.~F.~Li and Y.~L.~Zhou, Shifts of neutrino
oscillation parameters in reactor antineutrino experiments with
non-standard interactions, Nucl.\ Phys.\ B {\bf 888} (2014) 137
[arXiv:1408.6301 [hep-ph]]; and Y.~F.~Li, in section 3.2 of
Ref.~\cite{JUNO}.

\bibitem{Lisi2} F.~Capozzi, E.~Lisi and A.~Marrone,
Neutrino mass hierarchy and precision physics with medium-baseline
reactors: Impact of energy-scale and flux-shape uncertainties,
Phys.\ Rev.\ D {\bf 92} (2015) 093011 [arXiv:1508.01392 [hep-ph]].

\bibitem{MSW} L.~Wolfenstein, Neutrino oscillations in matter,
Phys.\ Rev.\ D {\bf 17} (1978) 2369.

\bibitem{MSW2} S.~P.~Mikheev and A.~Y.~Smirnov,
Resonance amplification of oscillations in matter and spectroscopy of
solar neutrinos, Sov.\ J.\ Nucl.\ Phys.\  {\bf 42} (1985) 913
[Yad.\ Fiz.\  {\bf 42} (1985) 1441].

\bibitem{Zhang} H.~Zhang and Z.~z.~Xing, Leptonic unitarity triangles
in matter, Eur.\ Phys.\ J.\ C {\bf 41} (2005) 143 [hep-ph/0411183].

\bibitem{XingZhu} Z.~z.~Xing and J.~y.~Zhu,
Matter-enhanced CP violation and Dirac unitarity triangles in a
low-energy medium-baseline neutrino oscillation experiment,
arXiv:1603.02002 [hep-ph].

\bibitem{Shrock} See, e.g., I.~Mocioiu and R.~Shrock,
Matter effects on neutrino oscillations in long baseline experiments,
Phys.\ Rev.\ D {\bf 62} (2000) 053017 [hep-ph/0002149].

\bibitem{PDG} K.~A.~Olive {\it et al.} [Particle Data Group
Collaboration], Review of particle physics, Chin.\ Phys.\ C {\bf 38}
(2014) 090001.

\bibitem{Fogli} F.~Capozzi, G.~L.~Fogli, E.~Lisi, A.~Marrone,
D.~Montanino and A.~Palazzo, Status of three-neutrino oscillation
parameters, circa 2013, Phys.\ Rev.\ D {\bf 89} (2014) 093018
[arXiv:1312.2878 [hep-ph]].

\bibitem{Fogli2} F.~Capozzi, E.~Lisi, A.~Marrone, D.~Montanino and
A.~Palazzo, Neutrino masses and mixings: Status of known and unknown
$3\nu$ parameters, arXiv:1601.07777 [hep-ph].

\bibitem{Valle} D.~V.~Forero, M.~Tortola and J.~W.~F.~Valle,
Neutrino oscillations refitted, Phys.\ Rev.\ D {\bf 90} (2014) 9,
093006 [arXiv:1405.7540 [hep-ph]].

\bibitem{GG} M.~C.~Gonzalez-Garcia, M.~Maltoni and T.~Schwetz,
Updated fit to three neutrino mixing: status of leptonic CP violation,
JHEP {\bf 1411} (2014) 052 [arXiv:1409.5439 [hep-ph]].

\bibitem{WangXing}
Y.~Wang and Z.~z.~Xing, Neutrino Masses and Flavor Oscillations,
arXiv:1504.06155 [hep-ph].

\bibitem{Li:2013zyd}
Y.~F.~Li, J.~Cao, Y.~Wang and L.~Zhan, Unambiguous Determination of
the Neutrino Mass Hierarchy Using Reactor Neutrinos, Phys.\ Rev.\ D
{\bf 88} (2013) 013008 [arXiv:1303.6733 [hep-ex]].

\bibitem{MHsens1}
X.~Qian, D.~A.~Dwyer, R.~D.~McKeown, P.~Vogel, W.~Wang and C.~Zhang,
Mass Hierarchy Resolution in Reactor Anti-neutrino Experiments:
Parameter Degeneracies and Detector Energy Response, Phys.\ Rev.\ D
{\bf 87} (2013) 033005 [arXiv:1208.1551 [physics.ins-det]].

\bibitem{MHsens2}
S.~F.~Ge, K.~Hagiwara, N.~Okamura and Y.~Takaesu, Determination of
mass hierarchy with medium baseline reactor neutrino experiments,
JHEP {\bf 1305} (2013) 131 [arXiv:1210.8141 [hep-ph]].

\bibitem{Ciuffoli:2013ep}
E.~Ciuffoli, J.~Evslin and X.~Zhang,
Neutrino mass hierarchy from nuclear reactor experiments,
Phys.\ Rev.\ D {\bf 88} (2013)  033017 [arXiv:1302.0624 [hep-ph]].

\bibitem{MHsens3}
M.~Blennow, P.~Coloma, P.~Huber and T.~Schwetz, Quantifying the
sensitivity of oscillation experiments to the neutrino mass
ordering, JHEP {\bf 1403} (2014) 028 [arXiv:1311.1822 [hep-ph]].

\bibitem{RENO50sens}
M.~Y.~Pac, Experimental Conditions for Determination of the Neutrino
Mass Hierarchy with Reactor Antineutrinos, Nucl.\ Phys.\ B {\bf 902}
(2016) 326 [arXiv:1508.01650 [hep-ex]].

\bibitem{DYB1}
F.~P.~An {\it et al.} [Daya Bay Collaboration],
Observation of electron-antineutrino disappearance at Daya Bay,
Phys.\ Rev.\ Lett.\  {\bf 108} (2012) 171803 [arXiv:1203.1669 [hep-ex]].

\bibitem{DYB2}
F.~P.~An {\it et al.} [Daya Bay Collaboration],
Improved Measurement of Electron Antineutrino Disappearance at Daya Bay,
Chin.\ Phys.\ C {\bf 37} (2013) 011001 [arXiv:1210.6327 [hep-ex]].

\bibitem{DYB3}
F.~P.~An {\it et al.} [Daya Bay Collaboration], Spectral measurement
of electron antineutrino oscillation amplitude and frequency at Daya
Bay, Phys.\ Rev.\ Lett.\  {\bf 112} (2014) 061801 [arXiv:1310.6732
[hep-ex]].

\bibitem{DYB4}
F.~P.~An {\it et al.} [Daya Bay Collaboration], New Measurement of
Antineutrino Oscillation with the Full Detector Configuration at
Daya Bay, Phys.\ Rev.\ Lett.\  {\bf 115} (2015) 111802
[arXiv:1505.03456 [hep-ex]].

\bibitem{ND1}
A.~B.~Balantekin {\it et al.}, Neutrino mass hierarchy determination and other physics
potential of medium-baseline reactor neutrino oscillation experiments, arXiv:1307.7419 [hep-ex].

\bibitem{ND2}
E.~Ciuffoli, J.~Evslin, Z.~Wang, C.~Yang, X.~Zhang and W.~Zhong,
Advantages of Multiple Detectors for the Neutrino Mass Hierarchy Determination at Reactor Experiments,
Phys.\ Rev.\ D {\bf 89} (2014) 073006 [arXiv:1308.0591 [hep-ph]].

\bibitem{ND3}
H.~Wang, L.~Zhan, Y.~F.~Li, G.~Cao and S.~Chen, Mass hierarchy sensitivity of medium baseline
reactor neutrino experiments with multiple detectors, arXiv:1602.04442 [physics.ins-det].

\end{thebibliography}
\end{document}